\documentstyle[prd,aps]{revtex}
\tightenlines
\input epsf.tex
\def\DESepsf(#1 width #2){\epsfxsize=#2 \epsfbox{#1}}
\newcommand{\be}{\begin{equation}}
\newcommand{\ee}{\end{equation}}
\newcommand{\bea}{\begin{eqnarray}}
\newcommand{\beas}{\begin{eqnarray*}}
\newcommand{\eea}{\end{eqnarray}}
\newcommand{\eeas}{\end{eqnarray*}} 
\newcommand{\ba}{\begin{array}}
\newcommand{\ea}{\end{array}}
\begin{document}

\draft
%@twocolumnfalse\endcsname
%\preprint{\vbox{
%\hbox{UMD-PP-00-0}}}

\title{Neutrinos, Large Extra Dimensions and Solar Neutrino
Puzzle\footnote{Invited talk presented by R. N. Mohapatra at the workshop
on
{\it Lepton Flavor Violation}, Honolulu, October (2000)}}

\author{
D. O. Caldwell$^1$\footnote{e-mail:caldwell@slac.stanford.edu},
R. N. Mohapatra$^2$\footnote{e-mail:rmohapat@physics.umd.edu},
and
S. J. Yellin$^{1}$\footnote{e-mail:yellin@slac.stanford.edu} 
}

\address{
$^1$ Department of Physics, University of California, Santa Barbara,
CA, 93106, USA.\\
$^2$ Department of
Physics, University of Maryland, College Park, MD, 20742, USA.}

\maketitle

\begin{abstract}
{A desirable feature of models
with large extra dimensions and a TeV range string scale is
the possibility to understand an ultralight sterile
neutrino needed for a simultaneous understanding of solar, atmospheric
and LSND results as a Kaluza-Klein state of a bulk neutrino.
How these oscillation data are understood in quantitative detail is still
not clear. We have recently suggested a new way
to understand the solar neutrino data in this framework by
a combination of vacuum and small angle MSW transition of $\nu_e$ to
sterile KK modes of the bulk neutrino. This mechanism can be embedded
into an economical extension of the standard model in the brane to
understand the atmospheric and LSND data. 
After a brief discussion of how
small neutrino masses arise naturally in extra dimensional models, we
review this new suggestion.}
\\[1ex] PACS: {14.60.Pq;
14.60.St; 11,10.Kk;} 
\end{abstract} 

\vskip0.5in

%%%%%%%%%%%%%%%%%%%%%%%%%%%%%%%%%%%%%%%%%%%%
\section{Introduction}
One of the important predictions of string theories is the existence
more than three space dimensions. For a long time, it was believed that 
these extra dimensions are small and are therefore practically
inconsequential as far as low energy physics is concerned. However, recent
progress in the understanding of the nonperturbative
aspects of string theories have opened up the possibility that some of
these extra dimensions could be large\cite{horava,nima} without
contradicting observations. In particular, models, where some of the extra
dimensions have sizes as large as a milli-meter and where the string
scale is  in the few TeV range have attracted a great deal of
phenomenological
attention in the past two years. The basic assumption of these models,
largely inspired by generic observations in string theories is that the
space time has a brane-bulk
structure, where the brane is the familiar (3+1)
dimensional space-time, with the standard model particles and forces
residing in it and bulk consists of all space dimensions where
gravity and other possible gauge singlet particles live. 
The main interest in these models is of
course due to the fact that the low string scale provides an opportunity
to test them using existing collider facilities.

A major challenge to these theories comes from the neutrino sector, the
first problem being how one understands the small
neutrino masses in a natural manner. The conventional seesaw\cite{seesaw}
explanation which is believed to provide the most satisfactory way to
understand this, requires that the new physics scale (or the scale of
$SU(2)_R\times U(1)_{B-L}$ symmetry) be around $10^{12}$ GeV or higher and
clearly does not work for these models. So one must look for 
alternative ways. The second problem is that if one considers only the
standard model group in the brane, operators such as $LH LH/M^*$ could be
induced by string theory in the low energy effective Lagrangian. For
TeV scale strings this would obviously lead to unacceptable neutrino
masses.

One mechanism suggested in Ref.\cite{dienes} is to postulate
the existence of one or more gauge singlet neutrinos, $\nu_B$ in the
bulk which couple to the lepton doublets in the brane. After
electroweak symmetry breaking this coupling can lead to neutrino Dirac
masses, which are suppressed by the ratio $M_*/M_{P\ell}$. This is
sufficient to explain small neutrino masses and owes its origin
to the large bulk volume that suppresses the effective Yukawa couplings of
the Kaluza-Klein (KK) models of the bulk neutrino to the brane fields. 
In this class of models, naturalness of small neutrino mass requires that
one must assume the existence of a global B-L symmetry in the theory,
since that will exclude the undesirable higher dimensional operators from
the theory.

An alternative possibility\cite{pere} is to consider the brane theory to
have an extended gauge symmetry which contains
B-L  symmetry as a subgroup, which will eliminate the higher dimensional
operators. This is perhaps more in the spirit of string
theories, which generally do not allow global symmetries. Phenomenological
considerations, however, require that the local $B-L$ scale and hence the
string scale be of order of $10^{10}$ GeV or so. The extra dimensions
in these models could of course be large.

Regardless of which path one chooses for understanding small neutrino
masses, a very desirable feature of these models is that if the size of
extra dimensions is of order of a milli-meter, the KK excitations of the
bulk neutrino have masses in the range of $10^{-3}$ eV, which is in the
range
needed for a unified understanding of oscillation data\cite{cald}.

In this paper, we will focus on TeV scale models and discuss their
implications. In this
class of models,
one gets for the Dirac masses of the familiar
neutrinos $\nu_{e,\mu,\tau}$ 
\begin{eqnarray}
m_{0} = \frac{h v_{wk} M_*}{M_{P\ell}}.
\end{eqnarray}
The righthanded neutrino in the Dirac mass is a mode of the bulk neutrino
$\nu_B$.
For $M_*\sim 10$ TeV, this leads to $m_{\nu}\simeq 10^{-4} h$ eV. It is
encouraging that this number is in the right range to be of interest in
the discussion of solar neutrino oscillation if the Yukawa coupling $h$
is appropriately chosen.

If one takes these models for the neutrinos seriously, some immediate
questions arise: first,
while the KK modes of the bulk neutrinos are candidates for sterile
neutrinos if the bulk radius is in the range of millimeters, are their small
masses natural? Secondly, how well do the models
of this type describe the observed oscillation data from solar,
atmospheric and LSND experiments ? It is clear that they provide a
tempting possibility to simultaneously explain all data since the needed
sterile neutrino could be one of the KK modes of the bulk neutrino, as
already emphasized.

We discuss the answer to the first
question in section 2. As far as the second question is concerned, several
recent papers have addressed this
issue\cite{dvali,barbieri,apl,lukas,sar}.
In particular, in Ref.\cite{apl} it has been shown that while the overall
features of the solar and atmospheric data can be accomodated in minimal
versions of these models with three bulk neutrinos, it is not possible to
simultaneously explain the LSND observation for the $\nu_{\mu}-\nu_e$
oscillation probability, and one must incorporate new physics in the brane.

On a phenomenological level, a first glance at the values of the parameters
of the model such as $m_0$ from Eq.(1) and $R^{-1}\sim 10^{-3}$ eV
suggests that perhaps one should seek a solution of the solar neutrino
data in these models using the small angle MSW mechanism\cite{dvali}.
However, present Super-Kamiokande recoil energy distribution seems to
disfavor such an interpretation, although any definitive conclusion should 
perhaps wait till more data accumulates\cite{concha}. In any case if the
present trend of the data near the higher energy region of the solar
neutrino spectrum from Super-Kamiokande persists,
it is likely to disfavor the small angle MSW solution and favor a
vacuum oscillation. We therefore pursue the possibility of vacuum
oscillation between the $\nu_e$ and $\nu_s$ to solve the solar neutrino
puzzle motivated by a model that leads to desired parameters in a natural
manner\cite{cmy}.

As is known from many discussions in literature,  chlorine results play
a pivotal role in deciding on the nature of the oscillation
solution to the solar neutrino puzzle. To fit water and the
Gallium\cite{N2000} data in conjunction with the Chlorine data, one must
``kill'' the $^7$Be neutrinos almost completely. In the vacuum solution,
it is in general hard to achieve this without simultaneously suppressing the pp
neutrinos. Therefore, the general tactic adopted is to suppress the $^7$Be
neutrinos as much as possible (but not quite to zero) and suppressing the
$^8$B neutrinos to less than 30\% or so. This, however, suppresses the
higher energy $^8$B contributions below the level observed by
Super-Kamiokande collaboration. In the case of oscillation to active
neutrinos, this can be fixed by including the neutral current scattering 
which contributes about 16\% of the charged current data. On the other
hand, this disfavors the vacuum oscillation to sterile neutrinos which do
not have a neutral current component. As we will show later\cite{cmy}, if
the
sterile neutrino is a bulk mode, its KK modes play a significant role in
resolving this problem due to the existence of MSW transitions of $\nu_e$ 
to these modes for values of extra dimension sizes of interest\cite{dvali}.

\section{Naturalness of ultra light sterile neutrinos in brane-bulk
models}
 Before getting to the detailed discussion of the model, we make some
general comments about the naturalness of the ultralightness of the bulk
neutrino. The bulk neutrino ``self mass'' terms are constrained by the
geometry of the bulk and could therefore under certain circumstances be
zero. If that happens, the only mass of the KK states of the $\nu_B$ will
arise from the kinetic energy terms such as $\bar{\nu}_B
\Gamma^I\partial_I\nu_B$, where $I= 5,6...$ and will be given by $n/R$
where $R$ is the radius of the extra dimensions. In such a situation, an
ultralight $\nu_{B,KK}$ arises naturally.

The key to naturalness of the ultralight bulk neutrino is the geometry
that forbids both Dirac and Majorana mass terms. Let us give a few
examples. In five dimensions, if we impose the $Z_2$ orbifold symmetry
($y\rightarrow -y$), then it follows that the Dirac mass vanishes. Now if
we impose lepton number symmetry in the brane, the Majorana mass vanishes,
leaving us with no mass term for the bulk neutrino in 5-dimensions.

Another interesting example is the 10-dimensional bulk, where the bulk
neutrino is a {\bf 16}-component spinor, which when reduced to
4-dimensions leads to eight 2-component spinors. The interesting point is
that for a {\bf 16}-dimensional spinor, one cannot write a Dirac or
Majorana mass term consistent with 10-dimensional Lorentz invariance.
In this case, there is no need for assuming lepton number to get an
ultralight sterile neutrino. A similar situation is also expected in six
dimensions if we choose the bulk neutrino to be a 4-component complex
chiral spinor.

%%%%%%%%%%%%%%%%%%%%%%%%%%%%%%%%%%%%%%%%%%%%%%%%%
\section{Basic ideas of the model and the mixing of bulk modes}

The model we will present will consist of the standard model in the brane
(suitably extended to include some additional Higgs fields)
and one bulk neutrino. The bulk may be five, six or higher dimensional; we
will assume that only one of those extra dimensions is large. The bulk
neutrino will be assumed to couple to $\nu_e$\footnote{In general, the
bulk neutrino will couple to all three species of brane neutrinos;
however, when the $\nu_{\mu,\tau}$ masses are much bigger than the mass
parameters in the $\nu_e$ and $\nu_B$ sector, one can decouple the
$\nu_{\mu,\tau}$ and restrict oneself to only the $\nu_e,\nu_B$ sector,
to analyse the effect of the bulk neutrino,as we do in the paper.}. Let us
begin our discussion by focusing on the $\nu_e$ in the brane and one
bulk neutrino. Obviously, the
fields that could propagate in the extra dimensions are chosen to be
gauge singlets.  Let us denote bulk neutrino by $\nu_B( x^{\mu}, y)$.
It has a five dimensional kinetic energy term and a coupling to the brane
field $L(x^{\mu})$ given by
\be
 {\cal L} = \kappa  \bar{L} H \nu_{BR}(x, y=0) + \int dy\
 \bar{\nu}_{BL}(x,y)\partial_5 \nu_{BR}(x,y) + h.c.,
 \label{l1}
 \ee 
where from the five dimensional kinetic energy, we have only kept the
5th component that contributes to the mass terms of the KK modes in the
brane; $H$ denotes the Higgs doublet, and 
$\kappa = h {M^*\over M_{P\ell}}$ the suppressed Yukawa coupling.
It is worth pointing out that this suppression is independent of the
number and radius
hierarchy of the extra dimensions, provided that our bulk neutrino
propagates in the whole bulk. For simplicity, we will
assume that there is only one extra dimension with radius of
compactification as large as a millimiter, 
and the rest with much smaller compactification radii. The smaller
dimensions will only contribute to the relationship
between the Planck and the string scale, but their
%(\ref{eq1}) but its 
KK excitations will be
very heavy and decouple from neutrino spectrum.
Thus, all the analysis could be done as in five dimensions.
 
A second point we wish to note is that we will include new physics in the
brane that will generate a Majorana mass matrix for the three standard
model neutrinos as follows:
\begin{eqnarray}
{\cal M}~=~\left(\begin{array}{ccc} \delta_{ee}  & \delta_{e\mu}
&\delta_{e\tau} \\ \delta_{e\mu} & \delta_{\mu\mu} & m_0 \\
\delta_{e\tau} & m_0 & \delta_{\tau\tau}\end{array}.\right)
\end{eqnarray}
The origin of this pattern of brane neutrino masses will be discussed
in a latter section. In this section let's discuss the mixing pattern of the
bulk neutrinos with the brane ones. For this we will assume that $m_0 \gg
\delta_{ij}$; as a result the $\nu_{\mu,\tau}$ in effect do not affect the
mixing between the bulk neutrino modes and the $\nu_e$.

To discuss the effect of the bulk modes, note that the first term in Eq.
(\ref{l1})  will be responsible for the neutrino mass once the Higgs field
develops its vacuum expectation. The induced Dirac 
mass parameter will be given by $m= \kappa v$, which for $M^*= 1$ TeV is
about $h\cdot 10^{-5}$ eV. Obviously this value  depends only linearly
on the fundamental scale. Larger values for $M^*$ will increase $m$
proportionally.  After introducing the expansion of the bulk field in
terms of the KK modes, the Dirac mass terms in  (\ref{l1}) could be written
as 
 \be 
   ({\nu}_{e} \nu_{0B} {\nu}'_{B,-} \nu'_{B,+})\left(\begin{array}{cccc}
\delta_{ee} & m &\sqrt{2} m & 0 \\ m &0 & 0 & 0\\ \sqrt{2} m & 0 
& 0 & \partial_5\\ 0 & 0 & \partial_5 &0
 \end{array}\right)\left(\begin{array}{c}\nu_e \\ \nu_{0B} \\
 \nu'_{B,-} \\ \nu'_{B,+}\end{array}\right),
 \label{m1}
 \ee
where our notation is as follows: $\nu'_B$ represents the KK excitations;
the off diagonal term  $\sqrt{2} m$  is actually an infinite row vector
of the form $\sqrt{2} m (1,1,\cdots)$.  The operator $\partial_5$ stands
for the diagonal and infinite KK mass matrix whose $n$-th entry is  given
by $n/R$. This notation was
introduced in~\cite{pere} to represent the infinite mass matrix in a
compact manner.

Using this short hand notation makes it easier to calculate the exact
 eigenvalues and the eigenstates of this mass matrix~\cite{apl}. 
Simple algebra yields the characteristic equation 
\be
 m_n =\delta_{ee}+ {\pi m^2\over \mu_0} \cot(\frac{\pi m_n}{ \mu_0}),
 \label{char1}
 \ee
 where $m_n$ is the mass 
eigenvalue~\cite{dienes,dvali}.  The equation for eigenstates is 
\begin{eqnarray}
\tilde \nu_n =\frac{1}{N_n}\left[ \nu_e + {m\over m_n}\nu_{0B}
+ {\sqrt{2} m 
\left(m_n\nu'_{B,-} + \partial_5 \nu'_{B,+}\right)\over m_n^2 
- \partial_5^2 }\right],
\label{nus}
\end{eqnarray}
%To simplify the discussion, let us first work in the limit of
%$\delta_{ee}=0$. This is quite justified since we expect that
%$\delta_{ee} \ll m$. In
%this limit, the eigenstates can be written
%as~\cite{apl}
% \be 
% \tilde \nu_{nL} = {1\over N_n} \left[ \nu_L + 
% {\sqrt{2} m \partial_5\over m_n^2 - \partial_5^2 }\, \nu'_{BL}\right],
% \label{nus}
% \ee
where the sum over the KK modes in the last term is implicit, and 
$N_n$ is the normalization factor given by
\be
N_n^2 = 1 + m^2\pi^2R^2 + {(m_n-\delta_{ee})^2\over m^2}.
\label{Nn}
\ee
% \be
%  N^2_n = {1\over\xi^2}\left(\lambda_n^2 + f(\xi)\right),
% \label{Nn}
%  \ee
%where $f(\xi)= \xi^2/2 + \pi^2 \xi^4/4$, with $\xi = {\sqrt{2} m \over
%\mu_0}$.
Using the expression (\ref{nus}), we can write down the weak eigenstate
 $\nu_L$  in terms of the massive modes as
 \be
 \nu_L = \sum_{n=0}^\infty {1\over N_n} \tilde \nu_{nL}.
 \label{nul}
 \ee
Thus, the weak eigenstate is actually 
a coherent superposition of
an infinite number of massive modes. Therefore, even for this single
flavour case, the time evolution of the mass-eigenstates involves in
principle all mass eigenstates and is very different from the simple
oscillatory behaviour familiar from the conventional two or three neutrino
case. The survival probability
depends strongly on the  parameter $\xi$, reflecting the
universal coupling of all the KK components of $\nu_B$ 
with $\nu_L$ in (\ref{l1}).
%It will depend on the value of $\xi$.
We will
be interested in the small $\xi$ region of the parameter space. In this
case it is easy to see that the coupling of the nth mode is given by
$\xi/n$. 

Note that in the limit of $\delta_{ee}=0$, the $\nu_e$ and $\nu_{B,0}$ are
two, two-component spinors that form a Dirac fermion with mass m. Once we
include the effect of $\delta_{ee}\neq 0$, they become Majorana fermions
with masses given by: $m_1\approx -\delta_{ee}/2 + m $ and $m_2\approx 
-\delta_{ee}/2 - m$, and they are maximally mixed; i.e., the two mass
eigenstates are $\nu_{1,2}\simeq \frac{\nu_e \pm \nu_{B,0}}{\sqrt{2}}$.
Thus as the $\nu_e$ produced in a weak interaction process evolves, it
oscillates to the $\nu_{B,0}$ state with an oscillation length $L\simeq
E/(2m\delta_{ee})$. Later on in this paper we will find that for natural
values of $m,\delta_{ee}$ in this model, $L$ is of order of the Sun-Earth
distance so that our model leads to vacuum oscillation of the solar
neutrinos. 

Furthermore since the $\nu_e$ also mixes with the KK modes of the bulk
neutrinos with a mass difference square of order $10^{-6}$ eV$^2$, there
is MSW resonance transition of $\nu_e$ to $\nu_{B,KK}$ modes. In our
discussion of the solar neutrino puzzle, we will take into account both
these effects.

In our model, we will also have $m_0$ in the brane neutrino mass matrix
which is of order eV so that we can account for LSND results. Atmospheric
neutrino oscillation is now purely $\nu_{\mu}$-$\nu_{\tau}$ oscillation.

\section{Solar neutrino data by a combination of vacuum and MSW
oscillations}
Let us now discuss how the solar neutrino data is explained within this
model. To discuss the MSW effect for the case of bulk neutrinos, we need
to include the matter effect in the diagonalization of the infinite
dimensional neutrino mass matrix.
The eigenvectors and eigenvalues  of this matrix can be found in the
presence of
matter, when the squared mass matrix, $M^2$, is replaced with
$M^2+2EH_1$, where $H_1$ is
 $\rho_e=\sqrt{2}G_F(n_e-0.5n_n)$ when acting on $\nu_e$,
and is zero on sterile neutrinos.  Define
\be
w_k={E\rho_e\over m_k\delta_{ee}} + \sqrt{1 + 
\left({E\rho_e\over m_k\delta_{ee}}\right)^2};
\label{wformula}
\ee
$w_k=1$ in vacuum.  The characteristic equation becomes
\be
m_k=w_k \delta_{ee} + {\pi m^2\over \mu_0} cot{\pi m_k\over\mu_0},
\label{characteristic5}
\ee
The eigenvectors are as in Eq. \ref{nus}, except the coefficients of
$\nu_{0B}$ and $\nu'_{B,-}$ 
acquire an additional factor $1/w_n$,
and the normalization becomes
$N_n^2 = 1 + {(1+{1\over w^2_n})\over 2}\left(\pi^2m^2R^2 +
{(m_n-w_n\delta_{ee})^2\over m^2}\right) - {(1-{1\over w^2_n})\over 2}
{(m_n-w_n\delta_{ee})\over m}$.

In fitting the solar data the simplest way to reconcile the
rates for the three classes of experiments is to ``kill'' the $^7$Be
neutrinos, reduce the $^8$B neutrinos by half and leave the pp neutrinos
alone. To achieve this in the VO case, one
may put a node of the survival probability function $P_{ee}$ around $0.86$
MeV. However, for an arbitrary node number, the oscillatory behaviour of
$P_{ee}$ before and after $0.86$ MeV cannot in general satisfy the other
two requirements mentioned above. If one uses the first node
to ``kill $^7$Be'', then for $^8$B neutrino energies the
$P_{ee}$ is close to one and not half as would be desirable; on the other
hand, if one uses one of the higher nodes (higher $\Delta m^2$), then pp
neutrinos get reduced.
The strategy generally employed is not to ``shoot'' for a node at the
precise $^7Be$ energy but rather somewhat away so that it reduces
$^7$Be to a value above zero. This requires that
one must reduce
%kill $^7$Be altogether but reduce it and reduce
the $^8$B neutrinos by much
more than 50\%, so one can fit Chlorine data. The water data then requires
an additional contribution, which,
in the case of active VO, is provided by the
neutral current cross section amounting to about 16\% of the charged
current one. Thus in a pure two-neutrino oscillation
picture, VO works for oscillation to active neutrinos but does not work
for active to sterile oscillation. It is here that the large extra
dimensions come to the rescue as we see below, thereby keeping the active
to sterile VO fit to solar neutrino data viable.

To see this note that in our model
both vacuum oscillations and MSW oscillations are
important. This is because
the lowest mass pair of neutrinos is split by a very small mass
difference, whereas the KK states have to be separated by
$<10^{-2}$ eV because of the limits from gravity experiments.
%when $\mu_0\approx 2\times 10^{-3}$ eV, $m_0 \approx 0.05\mu_0
We can then use the first node of $P_{ee}$ to
suppress the $^7$Be.  Going up in energy toward $^8$B neutrinos, the
survival probability, which in the VO case would have
risen to very near one, is suppressed by the
small-angle MSW transitions to the different KK excitations of the bulk
neutrino. This is the essence of our new way to fit the solar neutrino
observations\cite{cmy}.

To carry out the fit, we studied the time evolution of the $\nu_e$ state
using a program that evolved from one supplied by
W. Haxton\cite{Haxton}.  The
program was updated to use the solar model of BP98\cite{BP98} and modified
to do all neutrino transport within the
sun numerically.  For example, no adiabatic approximation was used.
Changes were also necessary for oscillations into sterile neutrinos and
to generalize beyond
the two-neutrino model.  Up to 16 neutrinos were allowed, but no more than
14 contribute for the solutions we considered.

\begin{figure}[htbp]
\epsfxsize=3in
\begin{center}
\leavevmode
\epsfbox{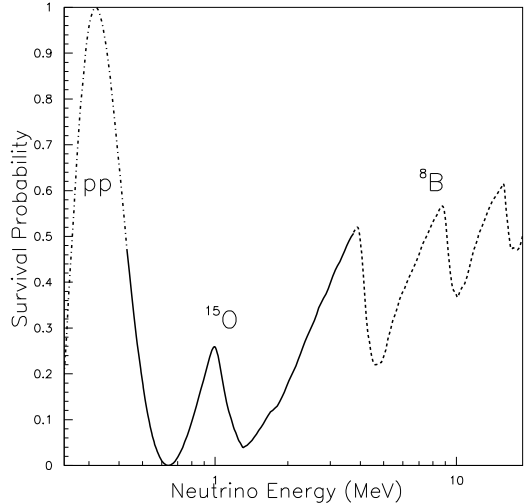}
\end{center}
\caption{Energy dependence of the $\nu_e$ survival probability when
$R=58\mu$m, $mR=0.0093$,
%$m=0.32\times 10^{-4}$ eV, and
$\delta_{ee}=0.84\times 10^{-7}$ eV.  The dot-dashed part of the curve
assumes the radial dependence in
the Sun for neutrinos from the pp reaction, the solid part assumes
$^{15}O$
radial dependence, and the dashed part assumes $^8B$ radial dependence.}
\label{fig:edep}
\end{figure}

For comparison with experimental results, tables of detector sensitivity
for the Chlorine and Gallium experiments were taken from Bahcall's web
site\cite{BP98}.
The Super-Kamiokande detector sensitivity was modeled
using \cite{Nakahata}, where the percent
resolution in the signal from Cerenkov light,
averaged over the detector for various total electron energies, is
provided. For details see \cite{cmy}.

Calculations of electron neutrino survival probability, averaged over the
response of detectors, were compared with measurements.  While theoretical
uncertainties in the solar model and detector response were included
in the computation of $\chi^2$ as
described in Ref. \cite{Fogli}, the measurement results given here include
only
experimental statistical and systematic errors added in quadrature.
The Chlorine survival probability, from Homestake\cite{Homestake}, is
$0.332\pm 0.030$.
Gallium results\cite{N2000} for SAGE, GALLEX and GNO were combined to give
a survival probability of $0.579\pm 0.039$.
The $5.5-20$ MeV 1117 day Super-K experimental survival
probability\cite{N2000} is $0.465\pm 0.015$.
The best fits were with $R\approx 58\mu m$, $mR$ around $0.0093$, and
$\delta_{ee}\sim 0.84\times 10^{-7}$ eV, corresponding to
$\delta m^2\sim 0.53\times 10^{-11}$ eV$^2$.
These parameters give average survival probabilities for Chlorine,
Gallium,
and water of %0.376, 0.512, and 0.463,  0.382, 0.533, 0.454,
0.386, 0.533, and 0.460,
respectively.  They give a $\nu_e$
survival probability whose energy dependence is shown in
Fig. \ref{fig:edep}.
For two-neutrino oscillations, the coupling between $\nu_e$
and the higher mass neutrino eigenstate is given by
$sin^22\theta$, whereas here the coupling between $\nu_e$
and the first KK excitation replaces $sin^22\theta$ by $4m^2R^2 =
0.00035$.

Vacuum oscillations between the lowest two mass eigenstates nearly
eliminate
electron neutrinos with energies of
.63MeV/(2n+1) for n = 0, 1, 2, ... .  Thus Fig. \ref{fig:edep}
shows nearly zero $\nu_e$
survival near 0.63 MeV, partly eliminating the $^7Be$
contribution at $0.862$ MeV, and giving a dip at the lowest neutrino
energy.
Increasing $\delta_{ee}$ moves the low energy dip to the right
%, which exacerbates the Gallium problem by moving the lowest energy dip
into Gallium's most sensitive pp energy range, making the fit worse.
Decreasing $\delta_{ee}$ increases Gallium,
but hurts the Chlorine fit by moving the higher energy
vacuum oscillation dip further to the left of the $^7Be$ peak.
MSW resonances start causing the third and fourth eigenstates to be
significantly occupied above $\sim 0.8$ MeV, the fifth and sixth
eigenstates
above $\sim 3.7$ MeV, the 7'th and 8'th above $\sim 8.6$ MeV, and the 9'th
and 10'th above $\sim 15.2$ MeV.
Fig. \ref{fig:edep} shows dips in survival
probability just above these energy thresholds.
The typical values of the survival probability within the
$^8B$ region ($\sim 6$ to $\sim 14$ MeV) are quite sensitive to the value
of
$mR$.  As can be seen from Eq. \ref{Nn}, higher $mR$
increases $1/N\approx m/m_n\approx mR/n$ for various
$n$, and thereby increases $\nu_e$ coupling to higher mass eigenstates,
strengthens MSW resonances, and lowers $\nu_e$ survival probability.

\begin{figure}[htbp]
\epsfxsize=3in
\begin{center}
\leavevmode
%\epsffile{skspectrum.eps}
\epsfbox{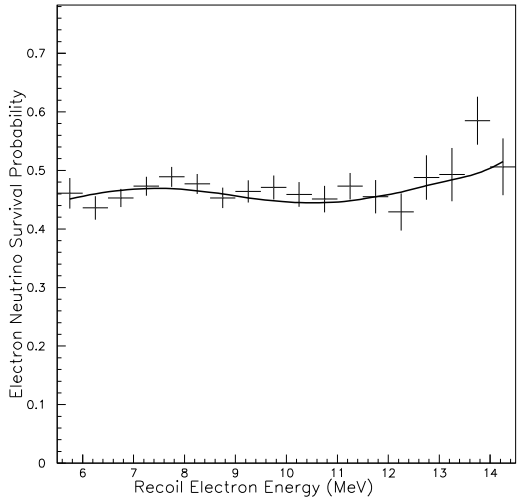}
\end{center}
\caption{Super-Kamiokande energy spectrum: measured [19]
preliminary results based on 1117 days (error bars)
and predicted (curve) for the same parameters as in Fig. 1.
The curve is not a fit to these data.}
\label{fig:skspect}
\end{figure}

The expected energy dependence of the $\nu_e$ survival probability is
compared with preliminary Super-K data\cite{N2000}
in Fig. \ref{fig:skspect}.  The uncertainties are statistical only.
The parameters used in making Fig. \ref{fig:skspect} were chosen to
provide
a good fit to the total rates only; they were not adjusted to fit
this spectrum.  But combining spectrum data with rates using the method
described in Ref. \cite{GHPV}, with numbers
supplied by Super-K\cite{Totsuka}, gives $\chi^2=15.6$ (``probability''
74\%)
for the spectrum predicted from the fit to total rates. It's
incorrect to calculate probability as if there were (number of data
points)
- (number of parameters) = 17 degrees of freedom, but if we do, $\chi^2$
corresponds to a ``probability'' of 55\%.

One may also seek fits with $\delta_{ee}$ constrained to be very small,
thereby eliminating vacuum oscillations.  The best such fit had
$\chi^2=5.5$
(``probability'' $14\%$).  The same parameters then used with the Super-K
spectrum gave $\chi^2=18.7$ (``probability'' $35\%$).

The seasonal effect was computed for a few points on the earth's orbit.
If $r$ is the distance between the earth and the sun,
\be
{r_0\over r} = 1 + \epsilon\,cos(\theta-\theta_0),
\label{eorbit}
\ee
where $r_0$ is one astronomical unit, $\epsilon=0.0167$ is the orbital
eccentricity, and $\theta-\theta_0\approx 2\pi(t-t_0)$, with $t$ in years and
$t_0\ =$ January 2, 4h 52m.  Table \ref{Seasonal}
shows very small seasonal variation.

To understand the mass
pattern used, consider
$L_e+L_{\mu}-L_{\tau}$ symmetry for neutrinos, with $L_e=1$ for $\nu_B$.
The allowed mass pattern for $(\nu_e, \nu_{\mu},
\nu_{\tau})$ is given by Eq. \ref{char1} with all $\delta$'s
(except $\delta_{e\tau,\mu\tau}$) set to zero. The remaining $\delta$
entries
arise after we turn on the symmetry breaking. For an explicit realization,
we augment the standard model by the singlet
charged Higgs $\eta^+$, $h^{++}$ which are blind with respect to lepton
number and $SU(2)_L$ triplet fields, $\Delta_{e,\mu,\tau}$,
with $Y=2$ which carry two negative units of lepton numbers
$L_{e,\mu,\tau}$, respectively. The Lagrangian involving these
fields consists of two parts: one ${\cal L}_0$ which is invariant under
$(L_e+L_{\mu}-L_{\tau})$ number and contains terms
($\eta L_{\mu}L_{\tau},\eta L_eL_{\tau},
h^{++}e^-_R\tau^-_R, h^{++}\mu^-_R\tau^-_R)$ and
$L_eL_e\Delta_e, L_{\mu}L_{\mu}\Delta_{\mu},
L_{\tau}L_{\tau}\Delta_{\tau}$
and a soft breaking term
${\cal L}_1 = h^{++}(\sum_{i= e,\mu,\tau} M_{ii}\Delta^2_i + M_{0\mu}
\Delta_{\mu}\Delta_{\tau}+M_{0\tau} \Delta_{e}\Delta_{\tau})
+ h.c.$
With these couplings, the neutrino Majorana masses arise from two-loop
effects (similar to the mechanism of ref.\cite{babu}), and we have
%the property
the $L_e+L_{\mu}-L_{\tau}$ violating entries $\delta_{ij} \sim c
~m_{e_i}m_{e_j}$. Using $\delta_{ee} \sim
10^{-8}$ eV, then $\delta_{\tau\tau} \sim 10^{-2}$, as would be required
to understand the atmospheric neutrino data.

Finally, models with bulk sterile neutrinos lead to new contributions to
low energy weak processes, which have been addressed in several
papers\cite{pospel}. In the domain of astrophysics, they lead to
new contributions to supernova energy loss, as well as to the energy
density in the early universe, which can influence the
evolution of the universe. Currently these issues are under
discussion\cite{george,barbieri,wu},
and if neutrino oscillation data favor these models, any
 cosmological constraints must be
addressed.  Note, however, that this model is less sensitive to these
issues than other fits because for VO $\Delta m^2$ is an order of
magnitude
smaller, for MSW $sin^22\theta$ is more than an order of magnitude
smaller,
and there is only a single KK tower starting from a very small
mass\cite{george}.

\begin{table}[h]
\caption{Predicted seasonal variations in $\nu_e$ fluxes, excluding
the $1/r^2$ variation.  The model assumed
$\mu_0=0.32\times 10^{-2}$ eV, $m_0=0.34\times 10^{-4}$ eV, and
$\delta_{ee}=0.78\times 10^{-7}$ eV.}
\vskip 2mm
\begin{center}
\begin{tabular}{|l|c|c|c|}
$\theta-\theta_0$ in eqn \ref{eorbit}& Chlorine & Gallium & Water\\ \hline
0 (January 2)    & 0.3787   &  0.5144 & 0.4635 \\
$\pm \pi/2$      & 0.3762   &  0.5121 & 0.4633 \\
$\pi$ (July 4)   & 0.3747   &  0.5082 & 0.4631 \\
\end{tabular}
\end{center}
\label{Seasonal}
\end{table}

The work of R.N.M. is supported by a grant from the National Science
Foundation No. PHY-9802551.  The work of D.O.C. and S.J.Y. is supported
by a grant from the Department Of Energy No. DE-FG03-91ER40618.
One of us (R. N. M.) would like to thank the organizers of the ``Lepton
Flavor Violation Workshop'' for kind hospitality and partial support.

 %%%%%%%%%%%%%%%%%%%%
%\vskip1em

\end{document}